\documentstyle[preprint,revtex]{aps}
\begin{document}
\draft
\begin{title}
\bf
Kondo Resonance in Transport Properties of \\ Double Barrier Structure.
\end{title}
\author{V.V.Ponomarenko}
\begin{instit} Frontier Research Program, RIKEN, Wako-Shi, Saitama
351-01, JAPAN.\\
Permanent address: A.F.Ioffe Physical Technical Institute,\\
194021, St. Petersburg, Russia.

\end{instit}

\date{\today}

\begin{abstract}
I consider the effect of the finite width of the resonant level on its
thermodynamics and tunneling transport properties in the
single electron charging regime. The finite width  of the levels
results from their delocalization with formation of a narrow band due to
their mutual overlap.
Making use of $1/N$ technique one can show that there is a fixed energy
level position at which the system undergoes the transition from
the Kondo regime (above this energy) to the paramagnetic regime (below it).
The  linear bias conductance calculated as a function of the chemical potential
turns out to be suppressed all around except for
the vicinity of the transition energy where it has an asymmetrical
resonant behavior.
\end{abstract}


\widetext

There is much interest in the study of  low temperature transport in
mesoscopic systems where new phenomena associated with quantum
coherence are invoked by the large Coulomb charging energy of a localized
constriction. These have been mostly discussed for the case of
tunneling through a
single spin-degenerate level specified with
 Anderson model \cite{Ng}.
Another way of manifesting a system with such a quantum correlation was
recently suggested for a conducting  granule
in the Coulomb blockade condition \cite{M}, \cite{Sh}.
In these systems, in particular,
the quasiclassical size of the granule results in the delocalization of the
states and the appearance of a continuum band instead of the separate levels.

In this paper I will consider the effect of the delocalization of the resonant
states on the Kondo
resonance phenomenon dealing with the problem of tunneling through
a double barrier (2D)  structure (DBS) located between two conductors.
This structure consists of the levels located at the impurity sites
between the barriers. The sites are ordered into a lattice, and the levels
of the neighbor cites have a small overlap, which leads to the appearance
of a narrow band of width  $E_G$.
First, one could discuss this system as a vertical quantum dot structure
where the tunneling layer is inside a constriction of radius $ R $.
This band should be considered continuous
if its level splitting $ \Delta_G \approx h v /R $ ( $v$ is the
speed of the band electron with energy near the bottom ) is less than
the inverse time $ \Gamma =\hbar/\tau $ of escape of the band electron
into the electrodes.
To make it more clear one could assume that this layer is infinite
and the effective tunneling area through the layer is only of $ R^2 $ size.
Then the
probability for the tunneling electron to escape from this area along the layer
could be estimated as relation between the squares
of appropriate surfaces
$ 2 \pi R v \tau $ and $ \pi R^2 $. This probability becomes negligible if
$ v\tau \ll R $.

The charging energy for accumulation of the charge
$ Q $ inside the barriers could be evaluated in a standard way \cite{AL}
as $ Q^2/2C $ and decreases as $ R^2 $ due to dependence of the capacitance
$ C(R) $ on the area of the contact. Since the charging  energy of a
non-equilibrium distributed charge should be more than the one in
equilibrium the criterion of a single-
electron charging of the layer could be formulated as $R<l_C$ where
$ e^2/ C(l_C) > \epsilon_b $, $ \epsilon_b $ is the bottom of the impurity
band.
To evaluate possible parameters of such a system
one could use as an example
the typical Metal-Oxide-Metal junction, where closely located conductors with
the
tunneling area of $100 nm^2$ form a condensator with $\approx 100K$ charging
energy. This value of the charging energy is substantially greater than
the band width which, due to the previous restriction according to $\Gamma $,
should be less than $0.1K$ for $10^2-10^3$ lattice sites.
The dependence on $R$ in this discussion could become irrelevant if
the phase coherence length of the  electrodes is less than $R$.
I could generalize the consideration if I will suppose that this length
satisfies the above two criteria instead of $R$.

Tunneling through the DBS is locally described
by the Hamiltonian
\begin{equation}
{\it H}= \sum_k \epsilon_{a,k} c^+_{ks,a} c_{ks,a} + \sum_q \epsilon_q
d^+_{qs}d_{qs} + \sum_{k,q,a,s} ( V_{a,q,k}d^+_{qs}c_{ks,a} +
V^*_{a,q,k}c^+_{ks,a}d_{qs} )  \label{1}
\end{equation}
with a restriction on the Hilbert space that only states of zero
$ |0> $ and one-electron $ |qs> $ occupation of the DBS band are permissible.
Here $ c_{ks,a}$ annihilates the electron of $a$ lead with spin $s$ in the
state of  wave vector $k=(k_{\perp },k_{\parallel})$ and of energy
$ \epsilon_{a,k}$. The components $k_{\perp },k_{\parallel}$ are perpendicular
and parallel to the layer, respectively. The operator $d_{qs}$ annihilates
an electron in the state localized inside the layer with
spin $s$, and wave vector $q$, and energy $ \epsilon_{q}$. An
expression for the tunneling matrix elements $ V_{a,q,k} $ could be generally
written making use of the tight binding model for the DBS band as
\begin{equation}
V_{a,q,k}={1\over {M L^{1/2}}} \sum_j T_{aj}(k_{\perp}) exp (iR_j q -
i\phi_{aj}(k_{\parallel}))
\label{2}
\end{equation}
Here $M$ is the effective number of the sites of the layer restricted
either by the geometry or by the coherence length. $L$ is the normalization
in the direction perpendicular to the layer, and $ T_{aj}$ is the modulus of
the corresponding tunneling amplitude.
 The vertex function may be introduced as
$ \Gamma_{a,\omega}(q_1,q_2)=\pi \sum_k \delta( \omega - \epsilon_k)
V^*_{a,q_1,k} V_{a,q_2,k}$
which is completely ample for our future needs.
Moreover, in the low temperature limit only the transitions between
the holes of the Fermi surface ($ \omega \approx 0 $)
and the electrons of the bottom of the layer band
are important. ( I will suppose that the bottom is reached at
$ \epsilon_{q=0}=\epsilon_b $.)
To calculate the function in this limit one could exploit either the jelly
model
for the leads or the tight binding model if all structure are commensurate.
Virtually, the result is the same in the above limit.
Indeed, taking $ \phi_{ja}=R_j k_{\parallel}$ one finds
\begin{equation}
\Gamma_{a,\omega}(q_1,q_2)=\pi  \sum_{Q,Q'} \delta_{k_{\parallel},q_1+Q}
\delta_{k_{\parallel},q_2+Q'} \int d\epsilon_{k_{\perp}}
\rho_{1D}(\epsilon_{k_{\perp}}) \delta(\omega-\epsilon_{k_{\perp}}
-\epsilon_{k_{\parallel}}) T^2_a(\epsilon_{k_{\perp}})\\
\simeq
\pi T_a^2 \rho_{a,1D}(0) \delta_{q_1,q_2}    \label{3}
\end{equation}
where $Q,Q'$ are the vectors of the reciprocal lattice either of the DBS or
of the leads. $T_{aj}$ is taken as a constant if the position of the $j$
DBS site corresponds to the position of the lead site ( and in the case of
the jelly model). Otherwise, it was put zero.
The second equation in (\ref{3})
is correct if the DBS lattice is less close than
the lattice of the leads. If this condition is not met, the
transitions between nonequal $q_1,q_2$ become permissible. I will not
discuss it specially, though, the consideration could be easy generalized
to account for them.

Conservation of the $q$ components in (\ref{3}) makes the tunneling resonant
and justifies the application of the $1/ N $ expansion technique
\cite{Bickers}.
In this case it is a $1/M $ expansion and it means the summing up of
the most divergent terms in $M$ of each order of expansion in $ \Gamma $.
I will restrict my consideration to leading order, giving summation of all
terms of $ ( \Gamma M)^n$.

Starting from the thermodynamics I will include the DBS contribution into
the partition function in terms of the self energy of the slave-boson
$ \Sigma_0(z) $ \cite{Bickers}. To leading order it takes the form
\begin{equation}
\frac{Z^{(1)}}{Z_{band}}=\sum_{q,s} e^{-\beta \epsilon_q} +
\int_{C} \frac{dz}{2\pi i} e^{-\beta z} \frac{1- \partial_z \Sigma^{(1)}_0(z)}
{z- \Sigma^{(1)}_0(z)}
\label{4}
\end{equation}
where $C$ encircles all the singularities of the integrand,
$\beta =1/T$ is the inverse temperature,   and
$\Sigma^{(1)}_0(z)$ is given by
\begin{equation}
\Sigma^{(1)}_0(z)= \sum_{a,s,q} \int^{\infty}_{-E_F} d\omega \Gamma_a
\frac{f(\omega)}{z-\epsilon_q+\omega}
\label{5}
\end{equation}
In the low temperature limit the Fermi-Dirac function $f(\omega)$ can be
approximate by the step function. Then $\Sigma^{(1)}_0$ for 2D band, which has
a constant density of states $ \rho_{2D}=m^*/(2\pi \hbar^2)$
near the edge, is
\begin{equation}
\Sigma^{(1)}_0(z)= \Delta \epsilon_b +P[E_Gln \frac{E_G+\epsilon_b-z}{E_G}
+(\epsilon_b-z)ln \frac{E_G+\epsilon_b-z}{\epsilon_b-z}]
\label{6}
\end{equation}
where $P=2(\Gamma/\pi)\rho_{2D}S \simeq(\Gamma/\pi)M/E_G,
\Gamma=\sum_a \Gamma_a, S$ is the square of the tunneling surface,
$\Delta \epsilon_b=P E_G(ln(E_G/E_F)-1)$.
$P$ is a number of the states lying inside the energy interval of width
$ \Gamma/\pi$.
In the above condition $P \gg 1$. This function (\ref{6}) coincides with the
one
of a single level Kondo
model for $|\epsilon_b-z| \gg E_G$. However, for small $|\epsilon_b-z| $ its
behavior is quite different. In particular,
the self-energy $\Sigma^{(1)}_0(z)$ converges to the finite negative value
$\Delta \epsilon_b $ at $\epsilon_b$ and acquires an imaginary part on
the real axis
above $\epsilon_b$ as $ Im \Sigma^{(1)}_0(x-i0)=\pi P(x-\epsilon_b) $.

This means that unless $\epsilon_b$ becomes lower then
$\Delta \epsilon_b $, the lowest
singularity of the integrand in (\ref{4}) is a simple pole
$ z=\epsilon_b -T_K$, which corresponds to the bound state of the slave-boson
with energy $T_K \equiv E_Gx$
\begin{equation}
\frac{ \bar{\epsilon_b}}{PE_G}=ln(1+x) +xln(1+x^{-1})+\frac{x}{P},
\ \ \ \ \ \
\bar{\epsilon_b}=\epsilon_b-P E_G(ln(E_G/E_F)-1)
\label{7}
\end{equation}
In the limit $\epsilon_b/|\Delta \epsilon_b| \ll 1$
the solution of (\ref{7}) loses its dependence on  $\epsilon_b$ and shows that
$T_K \simeq |\Delta \epsilon_b|$ when $E_F \gg E_G$.
In the more general case
$\bar{\epsilon_b}$ could reach the zero, where the transition into
a paramagnetic state of the DBS occurs. In that regime the low temperature
thermodynamics of the DBS is ruled out by the lowest states of its band.

The standard way of calculating \cite{Hewson}
the localized charge $n_G$, the charge
$\chi_C$ and the spin $\chi_s$ susceptibilities results, for the Kondo
regime behavior in
\begin{eqnarray}
F_{DBS}=-T ln \left[ \frac{Z^{(1)}}{Z_{band}} \right]
 =\epsilon_b -T_K ,\ \ \
n_G=\frac{\partial F_{DBS}}{\partial \epsilon_b}=1-Z_0 \nonumber \\
\chi_C=-\frac{P E_G Z_0^3}{T_K(E_G+T_K) } \ \ ,\ \
\chi_s=\frac{P E_G Z_0}{8T_K(E_G+T_K) } \label{8} \\
Z_0=\frac{1}{1-\partial_z \Sigma_0(z)|_{z=\epsilon_b-T_K}}=
\frac{1}{1+P ln[(E_G+T_K)/T_K]} \nonumber
\end{eqnarray}
The temperature dependence is dropped in Eq. (\ref{8}). It remains unimportant
while $PT/\Gamma \ll exp(T_K/T)$. These solutions (\ref{7}, \ref{8}) predict
the singular behavior of all thermodynamic characteristics at
$\bar{\epsilon_b}=0$. The function $n_G$
 goes to $1$ as $1/ln(\bar{\epsilon_b}/E_G)$.

In the paramagnetic regime the low temperature asymptotic for (\ref{4})
is
\begin{eqnarray}
\frac{Z^{(1)}}{Z_{band}}=\frac{M T}{E_G} e^{-\epsilon_b/T} a(\epsilon_b,T)
\label{9}  \\
a=1-\beta \int^{\infty}_0 dx \frac{\Gamma}{\pi} \frac{\bar{\epsilon_b} +Px}
{(\bar{\epsilon_b}+Px(1+ln(E_G/x)))^2+(\pi Px)^2}  \nonumber
\end{eqnarray}
If $-\bar{\epsilon_b}>TPln(E_G/T)$ the parameter $a$ is evaluated as
$a=1-\Gamma/(\pi \bar{\epsilon_b})$.
{}From  Eq.  (\ref{9}) one concludes
that $n_G=1$ and $\chi_s$ diverges as $1/T$.

Discussing the transport through this structure I will restrict consideration
to linear conductance. Then, the conductance can
be expressed in terms of the imaginary part
of the retarded Green function \cite{MW}
of the DBS electrons $G_q(\omega)$ labelled
with the quasi momentum $q$ and energy $\omega$ indices. Up to $T^2$
corrections
it is given by
\begin{equation}
\sigma_0=-\frac{2e^2}{\hbar} \frac{\Gamma_R \Gamma_L}{\Gamma^2}
P \sum_s \int d \epsilon_q Im G_q(\omega) |_{\omega=0}
\label{10}
\end{equation}
In this technique the full Green function of electron is calculated as a
convolution \cite{Bickers} of the slave-boson and the slave-fermion
Green functions along the appropriate contour in the complex energy plane.
For the imaginary part it is simplified to
\begin{equation}
Im G_q(\omega_+)=-\frac{1}{Z/Z_{band}} \int^{\infty}_{\infty}\frac{dx}{\pi}
\frac{Im \Sigma_0(x_+) Im \Sigma_q(x+\omega_+)}{|x-\Sigma_0(x_)|^2
|x+\omega-\epsilon_q -\Sigma_q(x+\omega)|^2} (1+e^{-\beta \omega})e^{-\beta x}
\label{11}
\end{equation}
Here $\Sigma_q(x_+)$ is the self energy of the slave fermion which is taken at
the real point $x_+=x+i0$ with the infinitely small imaginary shift.

In the Kondo regime the calculations straightforwardly
follow the ones of the single
level case. To leading order in the $1/M$ expansion, the  imaginary
parts of both the slave boson and the slave fermion Green functions
become strongly localized as $Z_0 \delta(\omega-T_K)$ and
$\delta(\omega-\epsilon_q)$, respectively, at low temperatures by the
factor $exp(-T/T_K)$. To obtain a non-zero $Im G_q(0)$ I should use the next
order expression for $ Im \Sigma_q(x)$ standing in the numerator of
(\ref{11}). This is written as
\begin{equation}
Im \Sigma_q(y_+)=Im \frac{\Gamma}{\pi} \int
\frac{dx}{y-x-\Sigma_0^{(1)}(y_+-x)}=-\Gamma Z_0 (1-f(y-\epsilon_b+T_K))
\label{12}
\end{equation}
for $y<\epsilon_b$. Finally, I obtain
\begin{equation}
Im G_q(\omega_+)=-\frac{\Gamma Z_0^2}{(\omega-\epsilon_q+\epsilon_b-T_K)^2},
\ \ \ \sigma_0=\frac{4e^2}{\hbar} \frac{\Gamma_R \Gamma_L}{\Gamma^2}
\frac{PZ_0^2 E_G}{T_K (T_K+E_G)}
\label{13}
\end{equation}
Taking $T_K$ from Eq.(\ref{7}) and $Z_0$ from Eq.(\ref{9}) one finds
\begin{equation}
\sigma_0=\frac{2e^2}{\hbar} \frac{\Gamma_R \Gamma_L}{\Gamma^2}
\frac{\Gamma}{\pi \bar{\epsilon_b} ln(E_G/\bar{\epsilon_b})}
\nonumber
\end{equation}
for small $T_K$. This shows that the resonance in the $\sigma_0$ dependence
on the bottom energy is reached at $\epsilon_b=\Delta \epsilon_b$, and it
decreases with increasing $\bar{\epsilon_b}$ more slowly than the
$1/\bar{\epsilon_b}^2$ tail of the one electron resonance. The width on this
side of the resonance is less than $\Gamma $ in the factor $1/ln(E_G/\Gamma)$.
Eq. (\ref{13}) is applicable only far from the resonance. Near the resonance
it could be corrected with the summation of the infinite series of the $1/M$
expansion as the NCA does.

In the paramagnetic phase, when $\epsilon_b$ lies below $\Delta \epsilon_b$,
the imaginary parts of the self energies give a contribution into the
conductance already in the leading order. However, this contribution is
$ \propto T$ since $\Sigma_0^{(1)}(y) \propto (y-\epsilon_b)$. The zero
temperature part of $\sigma_0$ appears if the imaginary part of $\Sigma_0$
standing in the numerator of (\ref{11}) is taken in the next order. Then
it is equal to
\begin{equation}
\Sigma_0^{(2)}(z)=P\frac{\Gamma}{\pi}\int_{-E_F}^0 d\epsilon {\large[}
\frac{1}{z+\epsilon-\epsilon_b-E_G}-\frac{1}{z+\epsilon-\epsilon_b}{\large]}
\int_0^{E_F}\frac{d\epsilon'}{z+\epsilon-\epsilon'-\Sigma_0^{(1)}(
z+\epsilon-\epsilon')}
\label{14}
\end{equation}
Extracting from here the imaginary part and collecting the leading terms
one could find
\begin{equation}
\sigma_0=\frac{8e^2}{\hbar}  \frac{\Gamma_R \Gamma_L}{\Gamma a
\bar{\epsilon_b}^2} P{\large(}\Gamma \frac{\phi(\bar{\epsilon_b})}{\pi}
+T {\large)}
\label{15}
\end{equation}
for $T<|\bar{\epsilon_b}|/(Pln\frac{E_G}{|\bar{\epsilon_b}|})$. The coefficient
$\phi$ is given by
\begin{equation}
\phi(\bar{\epsilon_b})=-\int_0^{E_F} \frac{d\epsilon}{\epsilon_b-\epsilon
-\Sigma_0^{(1)}(\epsilon_b-\epsilon)}
\simeq \frac{1}{P}lnln {\large(}\frac{E_GP}{|\bar{\epsilon_b}|} {\large)}
+ln {\large(}\frac{PE_G}{E_F+PE_G}{\large)}
\nonumber
\end{equation}
Equation (\ref{15}) describes the decay of the resonance below
$\epsilon_b=\Delta \epsilon_b$. It reveals that the square of the halfwidth
is larger than $\Gamma^2$ by a factor $P\phi$.

Finally, one could conclude that the delocalization of the resonant levels with
formation of a narrow band which is located inside the barrier in between
two electrodes, leads to the suppression of the Kondo conductance all around
except for a special position of the level energies at
$\epsilon_b=PE_G(ln(E_G/E_F)-1) $.
At this point the tunneling structure contributions to the thermodynamic
functions become singular, and the linear conductance has a resonant behavior.
These results are easy related to the known properties of the Coulomb blockade
transport through the granule \cite{AL}.
The above consideration describes the first steps
of the Coulomb staircase. Due to the lack of the electron-hole symmetry in the
excitation spectrum of the granule the position of the resonance is shifted
from $n_G=1/2$ \cite{M} to $n_G=1$,
and the form of the resonance undergoes asymmetrical
deformation \cite{Sh}.

I am grateful to D. Averin and K. K. Likharev for
elucidative discussions during my staying in
Stony Brook, and M. Stopa for his interest and kind help, and RIKEN ICO for
the hospitality. This
work was supported by the STA Fellowship (Japan) and by ONR grant
00014-93-1-0880.


\begin{references}
\bibitem{Bickers} N.E. Bickers, Rev. Mod. Phys. {\bf 59},845 (1987).
\bibitem{M} K.A.Matveev, Zh. Eksp. Teor. Phiz. {\bf 99}, 1598 (1991)
[Sov. Phys. JETP {\bf 72}, 892 (1991) ].
\bibitem{Hewson}  A. C. Hewson,
{\it The Kondon Problem to Heavy Fermions } (Cambridge
Univ. Press, Cambridge, England, 1993).
\bibitem{Ng} L.I. Glazman and M. E. Raikh, Pis'ma  Zh. Eksp. Teor. Phiz.
{\bf 47}, 378 (1988) [ JETP Lett. {\bf 47}, 452 (1988) ];
Y. Meir, N.S. Wingreen and P.A. Lee,
Phys. Rev. Lett. {\bf 70}, 2601 (1993);
T.K. Ng , Phys. Rev. Lett. {\bf 70}, 1768 (1993).
\bibitem{Sh} H. Schoeller and G. Schon,
Phys. Rev. B {\bf 50}, 18436  (1994).
\bibitem{AL}  D. Averin and K. K. Likharev,
in {\it Mesoscopic Phenomena in Solids },
edited by B. L. Altshuller {\it et al } (Elsevier, Amsterdam, 1991), p.173.

\bibitem{MW}L.I. Chen and C.S. Ting, Phys. Rev. B {\bf 41}, 8533  (1990);
Y. Meir and N.S. Wingreen,
Phys. Rev. Lett. {\bf 68}, 2512 (1992).

\end{references}
\end{document}